\documentclass[fleqn,10pt]{wlscirep}
\usepackage{microtype}
\usepackage{xcolor}
\usepackage[utf8]{inputenc}
\usepackage[T1]{fontenc}
\usepackage{caption}
\title{Geometric closure of classical nucleation theory for magnetic-field-controlled nanoparticle size across magnetic classes}

\author[1]{Yazeed Tawalbeh}
\author[1,2,*]{Mauro Fernandes Pereira}
\affil[1]{Department of Physics, Khalifa University of Science and Technology, 127788 Abu Dhabi, United Arab Emirates}
\affil[2]{Institute of Physics, Czech Academy of Sciences, 18221 Prague, Czech Republic}

\affil[*]{mauro.pereira@ku.ac.ae}
\begin{abstract}
Controlling nanoparticle size during synthesis remains a central challenge in nanoscience, particularly in systems where external magnetic fields are used as continuous control parameters. Existing descriptions of magnetic-field-assisted nucleation are typically material-specific or rely on computationally intensive atomistic methods.
Here, we reformulate classical nucleation theory as a geometrically closed thermodynamic framework by introducing a sphere-packing representation of atomic assembly. This construction establishes a direct link between discrete atomic structure and continuum free-energy contributions under applied magnetic fields, yielding a field-driven evolution equation for the critical nucleus size.
The resulting theory provides a unified description of nanoparticle nucleation across superparamagnetic, paramagnetic, and diamagnetic systems within a single formalism. It quantitatively reproduces previously unresolved experimental observations for magnetite and nickel nanoparticles, namely the systematic reduction of mean particle size and narrowing of size distributions with increasing magnetic field. In the diamagnetic limit, the framework recovers our earlier analytical susceptibility-based description of silver nanoparticles, in which the field-dependent critical radius is governed by the induced-magnetization contribution to the nucleation free energy.
Beyond modeling the reduction of mean particle size with increasing magnetic field, the framework reveals that the narrowing of size distributions emerges naturally from the curvature of the field-modified free-energy landscape. These results establish magnetic-field-assisted nucleation as a geometrically constrained thermodynamic process, providing a computationally efficient route for controlling nanoparticle size across distinct magnetic material classes.
\end{abstract}
\begin{document}

\flushbottom
\maketitle

\thispagestyle{empty}

    Nanoparticles (NPs) are generally defined as structures with characteristic dimensions below approximately 100 nm. At these reduced length scales, materials exhibit physical and chemical properties that differ markedly from their bulk counterparts. In particular, the functional performance of NPs is governed by their size and morphology, which determine the surface-to-volume ratio, electronic structure, and interfacial behavior. Precise control over these structural parameters is therefore a fundamental requirement to enable predictable functionality across nanoscale systems. By tailoring size and morphology, NPs can be engineered for a wide range of applications, including nanoelectronics, catalytic systems for chemical processing and energy, sensing and biomedical platforms, such as imaging agents, tumor detection tools, and controlled therapeutic delivery, as well as quantum devices, advanced construction materials, biofuel stimulants, and food-related technologies \cite{henini2011handbook, Razeghi2010, MISHRA2018631, D4MA90092H, Fabrizio12, Dong:20, D1MA00538C}.\\

Beyond their broad technological relevance, NPs are emerging as key building blocks in photonics and optoelectronics. In particular, they enable functionalities that are not readily accessible in bulk silicon platforms, thus extending the capabilities of silicon-based technologies \cite{Zafar2025-dl,Zafarpol4,Zafarreview,Zafarpol6,zafar2023band,zafar2023compact,ZafarIEEEAccess,Zafar:22}. Furthermore, NPs can act as versatile quantum emitters \cite{Uppu2021}, which underpin applications in quantum metrology and information science, where phenomena such as quantum squeezing enable enhanced measurement precision \cite{11142662}. In this context, NPs also provide a materials-efficient alternative to more complex architectures, such as superlattices and multiple quantum wells, for a wide range of optoelectronic applications \cite{Pereira2020-ii, 10.1117/1.JNP.11.046022,ma11010002,Vaks2022-uk}.\\

Despite these advances, achieving reliable control over NP size during synthesis remains a central and unresolved challenge in nanoscience. Because key properties, including catalytic activity, optical response, magnetic behavior, and electronic structure, are highly size-dependent, even small variations in the NP radius can lead to substantial changes in performance. Among the external parameters explored to control NP formation, magnetic fields have emerged as a particularly powerful and experimentally accessible tool, offering a non-invasive and continuously tunable means of influencing nucleation. A growing body of experimental evidence shows that applied magnetic fields can significantly modify the size, morphology, and size distributions of NPs across superparamagnetic, paramagnetic, and diamagnetic systems \cite{ma2022using, kthiri2021novel, li2025effects, ualkhanova2019influence, luo2015strong, kim2014situ}. However, these observations still lack a unified theoretical description, preventing the identification of general scaling relations between magnetic field strength, the free-energy landscape, and the critical nucleus size. Existing approaches are typically material-specific, limited to particular magnetic regimes, or rely on computationally intensive atomistic simulations. For example, our earlier susceptibility-only analytical treatments are restricted to specific regimes \cite{Tawalbeh2025controlling,Tawalbeh2026}, whereas density-functional-theory studies can resolve detailed cluster or NP energetics during nucleation and growth but are generally limited to specific material systems and atomistic structures on the order of hundreds of atoms \cite{Saidi2015PtMoS2, Wang2017PtAl2O3, Baek2025QuasicrystalDFT}. By contrast, the present thermodynamic formulation evolves the critical radius directly and can be applied efficiently to experimentally relevant NP size ranges beyond the practical atom-count limits of explicit DFT simulations.\\

In this work, we reformulate classical nucleation theory as a geometrically closed thermodynamic framework for magnetic-field-driven NP nucleation. By introducing a sphere-packing representation of atomic assembly, we establish a direct link between discrete atomic structure and continuum free-energy contributions, yielding a field-driven evolution equation for the critical nucleus size. The resulting framework provides a unified treatment of superparamagnetic, paramagnetic, and diamagnetic systems, while embedding our earlier regime-specific analytical descriptions \cite{Tawalbeh2025controlling, Tawalbeh2026} as limiting cases of a broader experimentally validated theory.\\

\begin{figure}[!ht]
    \centering
    \includegraphics[width=1\linewidth]{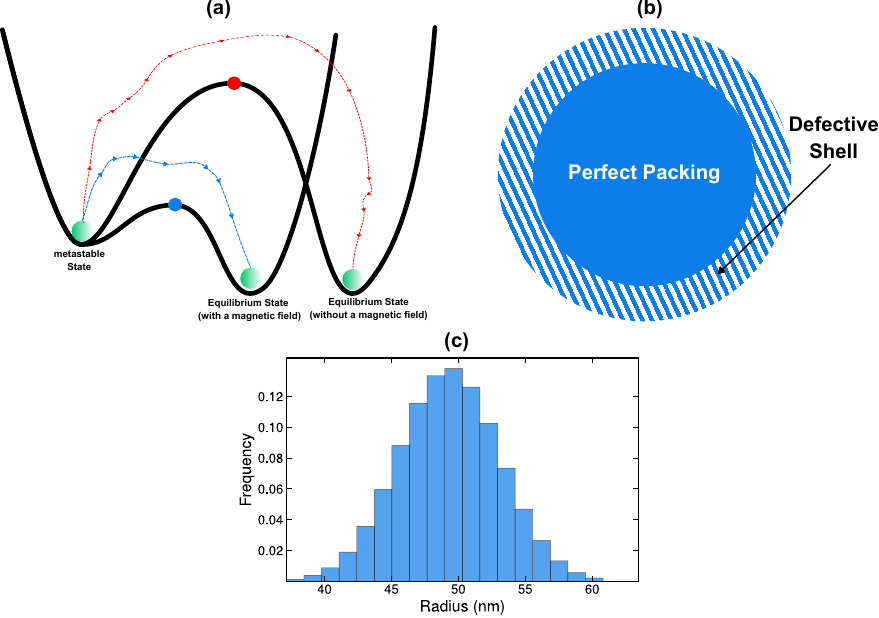}
    \caption{(a) Schematic illustration of the magnetic-field effect on the free-energy landscape. {The thin red and blue curves are guide lines representing the free-energy barrier along the nucleation coordinate. The red curve corresponds to the absence of a magnetic field, while the blue curve corresponds to the presence of a magnetic field, illustrating that the applied field lowers the barrier associated with critical-nucleus formation.} (b) Sphere-packing representation of a NP used in this work. (c) Model-generated size distribution for Ni NPs used in GaN nanowire growth \cite{kim2014situ} at the highest applied field in that dataset, $B=0.818$ T, chosen to illustrate the predicted distribution under the strongest field-induced size reduction.}
    \label{fig:1}
\end{figure}

The model represents each nucleus as a sphere-packed assembly consisting of a densely packed core surrounded by a defective shell of reduced packing efficiency, as illustrated in Fig. \ref{fig:1}b. This construction closes classical nucleation theory geometrically by mapping discrete atomic packing onto continuum free-energy contributions. Within this framework, the applied magnetic field modifies the nucleation free-energy landscape, lowering the barrier and shifting the critical nucleus size, as illustrated schematically in Fig. \ref{fig:1}a. The resulting field-dependent size distribution predicted by the model is shown for Nickel NPs in Fig. \ref{fig:1}c.

\section*{Results and Discussion}
In this work, we validate the generalized thermodynamic framework using experimental datasets spanning different magnetic classes \cite{Tawalbeh2025controlling, Tawalbeh2026} by establishing a connection between the magnetic field and the NP size that results under different magnetic configurations. We show that the framework successfully captures previously unexplained behavior in superparamagnetic magnetite NPs \cite{ma2022using}, paramagnetic nickel NPs used for carbon nanofiber \cite{luo2015strong} and gallium nitride (GaN) nanowire growth \cite{kim2014situ}, and diamagnetic silver NPs synthesized using different methods \cite{kthiri2021novel, li2025effects}.

We introduce the work of formation $\Delta F$ as the free-energy barrier associated with forming a post-critical nucleus from the metastable parent phase. In CNT, the critical nucleus corresponds to the stationary point at the top of this barrier along the nucleation coordinate, rather than to a stable equilibrium minimum. The total free energy is decomposed into four physically distinct contributions to separate the classical nucleation terms from the magnetic-field-dependent terms. The surface and bulk terms are the standard classical nucleation contributions, whereas the induced-magnetization and permanent-moment entropy terms account for the two magnetic responses needed to treat diamagnetic, paramagnetic, and superparamagnetic systems within one framework. The latter entropy term is absent in the susceptibility-only analytical limit \cite{Tawalbeh2025controlling,Tawalbeh2026}, which is why that earlier treatment cannot generally describe magnetic NPs with field-alignable moments. The resulting expression is

\begin{equation}
\Delta F(r,B)=
\underbrace{4\pi r^{2}\gamma}_{\text{surface formation}}
\;-\;
\underbrace{\,n(x)\,\Delta\mu}_{\text{bulk driving}}
\;-\;
\underbrace{ n(x)\Theta(\sigma)\frac{V_ {a}}{2}\frac{3}{\mu_0}\frac{|\chi_m|}{(3+\chi_m)} {B}^2}_{\text{induced magnetization}}
\;-\;
\underbrace{k_{B}T\,\ln\!\left[\frac{\sinh(\xi)}{\xi}\right]}_{\text{permanent moment}}
\label{eq:main}
\end{equation}

Here $x = r/a$, where $a$ is the effective atomic radius defined from the atomic volume of the material (see Methods). The surface formation term represents the energetic cost of creating a new interface, where $\gamma$ is the surface free energy. The bulk driving term accounts for the thermodynamic driving force through the chemical potential difference $\Delta \mu$, with $n(x)$ denoting the number of atoms in the NP. The induced magnetization term captures the response of the system to the applied magnetic field, where $V_a$ is the volume of the building unit and $\chi_m$ is the magnetic susceptibility. The function $\Theta(\sigma)$ accounts for the angular distribution of the induced magnetic moments relative to the field. It assigns a Gaussian weight to $\cos(\theta)$, where $\theta$ represents the angle between the induced magnetic moment and the magnetic field. The Gaussian has standard deviation $\sigma$ centered at $\theta = 0$ for paramagnetic systems and $\theta = \pi$ for diamagnetic systems. The final term represents the contribution of permanent magnetic moments, where  $\xi = mB/k_BT$, with $m = m_0 n(x)$, which is proportional to the size of the NP. Thus, the orientational degree of freedom is assigned to one effective collective NP moment rather than to $n(x)$ independent microscopic moments. We note that the contribution of the entropic term in the small $\xi$ limit becomes quadratic; however, it remains unrelated to the induced magnetization term and does not result in double counting the energy, as the energy comes from two different sources. A detailed derivation of all terms in Eq. (1), including the full distinction between the susceptibility-response contribution and the orientational-alignment contribution, is provided in the Supplementary Information. Ref.~\cite{VanVleck1932} makes the same basic distinction.

{We represent the number of atoms inside an NP as a densely packed core surrounded by a defective shell, as shown in Fig. \ref{fig:1}b. The expression of $n(x)$ is}
\begin{equation}
    n(x) = \phi_b(x-\delta)^3 + \phi_d[x^3 - (x-\delta)^3]
\end{equation}
{$\phi_b = \frac{\pi}{3\sqrt{2}}$ is the optimal close-packing fraction, and $\phi_d=0.517$ is the imperfect surface-packing fraction fitted from sphere-packing data \cite{pack}. $\delta$ is the thickness of the defective, imperfectly packed shell. The defective shell is included because surface atoms are undercoordinated and are most directly exposed to external forces from the nucleation medium, including solvent, ligand, ionic, and field-mediated interactions. Therefore, the surface region is not expected to maintain the same packing efficiency as the bulk-like interior. This picture is consistent with experimentally observed crystalline-core/amorphous-shell NP structures, where the interior remains comparatively ordered while the surface shell is structurally disordered \cite{Lu2016BlackTiO2,Kim2020Ni2PHo2O3}. The present mathematical model aims to capture this core-shell packing contrast through a dense core and a reduced-packing defective shell, thereby linking the discrete atomic count to the continuum free-energy terms. The reduced shell packing fraction is fitted from sphere-packing data to represent the effective packing loss of the surface layer. The model reduces to the densely packed limit when the defective shell vanishes, $\delta\to0$, or when $\phi_d=\phi_b$, giving $n(x)=\phi_b x^3$. For large particles with $x\gg\delta$, the shell contribution becomes relatively small and the model also approaches the densely packed bulk limit.}

In order to obtain the radius field relationship, we define the critical value of $\Delta F$ through a stationarity condition as $f(x,B)\equiv\frac{\partial\Delta F}{\partial x}=0$. This condition defines a one-dimensional critical manifold in the $(x,B)$ space. 

Rather than solving the stationarity condition independently for each value of the magnetic field, we reformulate the problem as a field-driven evolution along the critical manifold. {By differentiating $f(x(B),B)$, we track how the critical nucleus size, defined by the barrier stationarity condition, changes as $B \rightarrow B + dB$ and obtain an explicit differential equation for $dx/dB$}

\begin{equation}
{
\frac{dx}{dB}=
\frac{
n'(x)\left[
2KB+m_0 L(\xi)+\frac{n(x)m_0^2B}{k_BT}\left(-\mathrm{csch}^2\xi+\frac{1}{\xi^2}\right)
\right]
}{
8\pi\gamma a^2-n''(x)\left[\Delta\mu+KB^2\right]-n''(x)m_0B L(\xi)-\frac{\left[n'(x)\right]^2m_0^2B^2}{k_BT}\left(-\mathrm{csch}^2\xi+\frac{1}{\xi^2}\right)
}
}
\label{eq:dxdB}
\end{equation}
with $
K
=
\Theta(\sigma)\,
\frac{V_a}{2}\,
\frac{3}{\mu_0}\,
\frac{|\chi_m|}{3+\chi_m}
$, $L(\xi)\equiv \coth\xi-\frac{1}{\xi}$, {$\xi=n(x)m_0B/(k_BT)$, and $- \mathrm{csch}^2\xi+1/\xi^2=dL/d\xi$.}

We solve Eq. \eqref{eq:dxdB} numerically (at $\sigma = 0$) as an initial value problem with the initial condition $x(B=0)=x_0$ where $x_0$ is the peak of the most frequent radius inferred from the experiments. The parameter set $\boldsymbol{p}=\{\Delta \mu, \gamma, m_0,\delta\}$  is used as the least squares fit parameters varied iteratively to find the best fit parameter set $\boldsymbol{p}^{\ast}$. We resolve the equation again using different initial conditions that correspond to the boundaries of the experimental size distribution to obtain a theoretical Gaussian size distribution that we use as a probability density function, along with rejection sampling to generate size histograms that can be compared with experiments, as shown in Fig. \ref{fig:3}.

This equation defines a continuous trajectory of the critical nucleus size as a function of the applied magnetic field. {In classical nucleation theory, the critical nucleus represents the threshold size at which a cluster becomes thermodynamically favored to grow rather than dissolve. This threshold size sets the initial post-critical length scale from which later particle evolution proceeds. Therefore, comparison with experimentally measured NP radii is meaningful when post-nucleation processes, including subsequent growth, coalescence, aggregation, Ostwald ripening, and transport, either remain comparable across a given experimental series or do not reverse the field-dependent size hierarchy established during nucleation. Under this assumption, the measured particle radius is interpreted as the observable post-critical outcome of a field-selected critical seed, allowing the model to predict the magnetic-field dependence of the final radius.}

\subsection*{General Case: Magnetite and Nickel NPs}
\begin{figure}[!ht]
    \centering
    \includegraphics[width=1\linewidth]{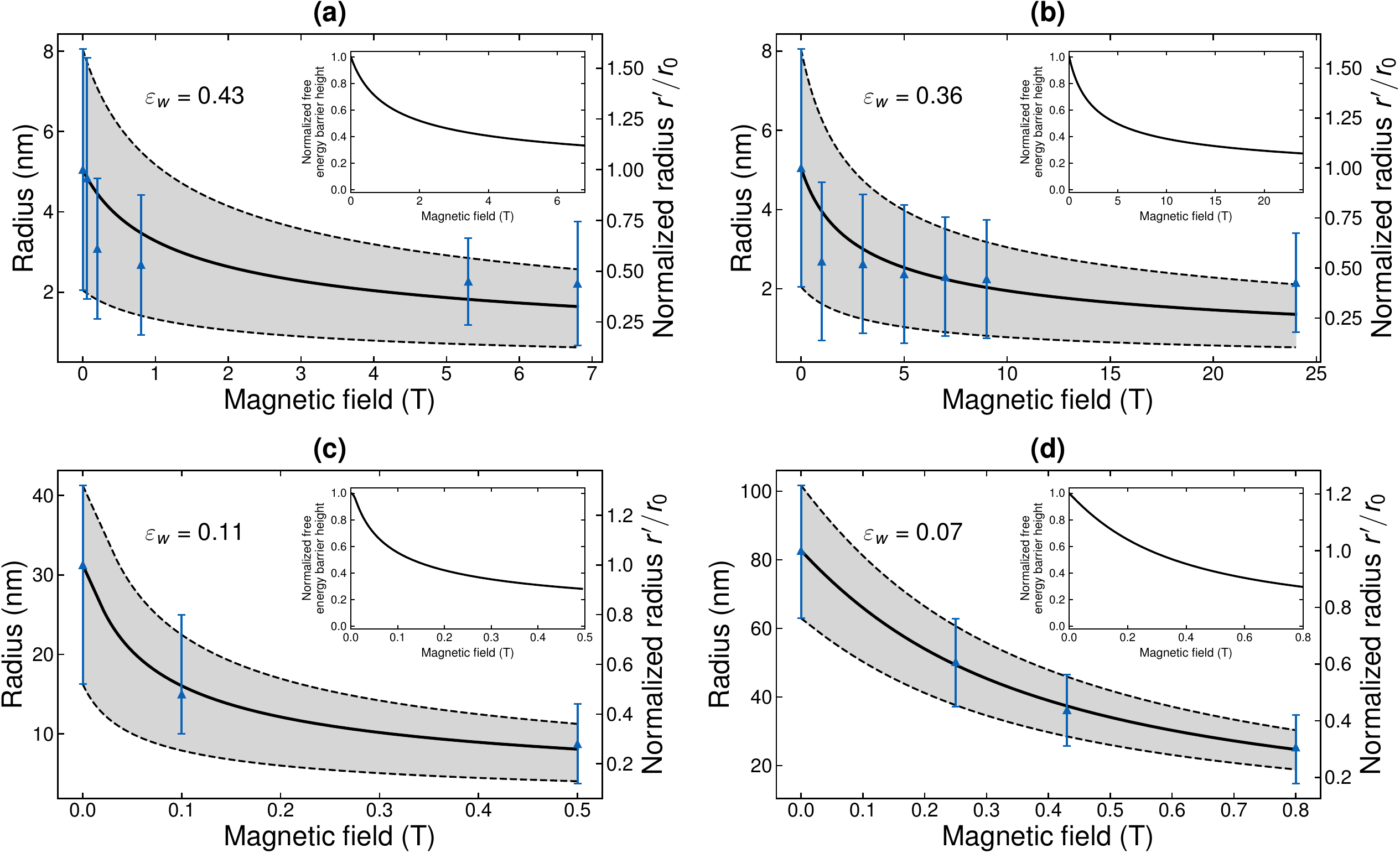}
    \caption{The radius–field relation for Magnetite NPs \cite{ma2022using} in (a) the gradient configuration and (b) the homogeneous configuration, and for Nickel NPs used in the synthesis of (c) carbon nanofibers \cite{luo2015strong} and (d) GaN nanowires \cite{kim2014situ}. The blue markers are experimental points; the bars represent the size distribution width, not to be confused with measurement errors. The black line represents the solution of Eq.\eqref{eq:dxdB} for $x(B=0)=x_0$, the dashed lines are the solutions at the experimental bounds. The shaded region represents the possible sizes generated by the theory, and the insets showcase the effect of the magnetic field on the free-energy barrier height. {The distribution-width-normalized RMS deviation, $\varepsilon_w=[N^{-1}\sum_i((r_i^{\mathrm{th}}-r_i^{\mathrm{exp}})/w_i)^2]^{1/2}$, where $w_i$ is the plotted radius distribution width, is given by $\varepsilon_w=$ 0.43, 0.36, 0.11, and 0.07 for panels (a)--(d), respectively.}}
    \label{fig:2}
\end{figure}

\begin{figure}[!ht]
    \centering
    \includegraphics[width=1\linewidth]{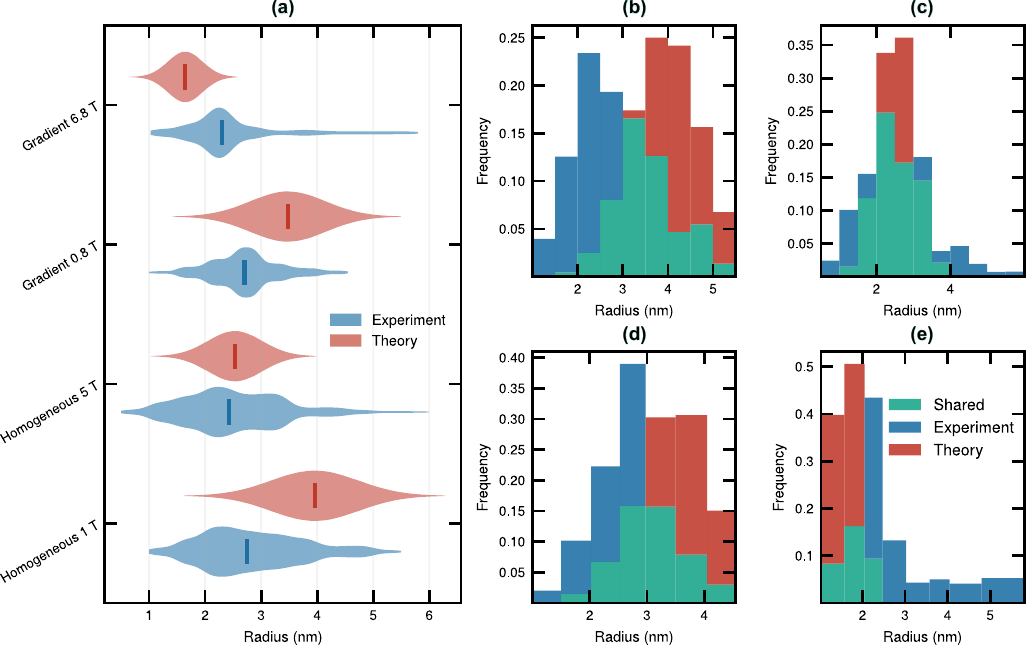}
    \caption{(a) A violin plot comparing the experimental and theoretical size distributions of magnetite NPs \cite{ma2022using} at selected magnetic-field values. {The blue distributions correspond to the experimental data, while the red distributions correspond to the model-generated distributions. The vertical bars indicate the central radius of each distribution. In the homogeneous-field configuration, the theory predicts the field-induced shift toward smaller radii as the field increases from 1 T to 5 T. Panel (b) shows the homogeneous 1 T case, where the theoretical distribution is shifted toward larger radii than the experimental distribution but retains partial overlap with the experimental range. Panel (c) shows the homogeneous 5 T case, where the theoretical and experimental distributions overlap more strongly and are both concentrated near smaller radii. Panels (d) and (e) show the gradient-field cases at 0.8 T and 6.8 T, respectively. At 0.8 T, the theory captures the general radius range but remains shifted toward larger radii, whereas at 6.8 T the model predicts a compact small-radius distribution while the experimental data retain a broader high-radius tail. The green regions in panels (b)--(e) denote the shared overlap between theory and experiment. Overall, Fig. 3 shows that the model captures the qualitative narrowing and field-driven reduction of the magnetite NP size distributions, with larger deviations in the gradient-field cases where spatial field nonuniformity and irregular particle morphology are expected to be more important.}}
    \label{fig:3}
\end{figure}

To compare the theoretical predictions of our model with experimental datasets, we use datasets for $\text{Fe}_3\text{O}_4$ Magnetite NPs \cite{ma2022using} and Nickel NPs \cite{luo2015strong, kim2014situ} reported in the literature. The Magnetite NPs are superparamagnetic, although Magnetite is ferromagnetic in bulk and their data are reported in two different configurations. The first is the homogeneous configuration in which Magnetite NPs are prepared in a chamber at {$200~^\circ\text{C}$} where the magnetic field is uniform throughout the chamber. The second configuration is the gradient configuration in which Magnetite NPs are prepared in a chamber under the same temperature but with a non-uniform magnetic field inside, such that different regions inside the chamber are subject to different magnetic fields. Figs. \ref{fig:2}a and \ref{fig:2}b showcase the radius-field relations for the gradient and homogeneous configurations, respectively. Fig. \ref{fig:3} shows an overall comparison between the size distributions obtained using our theory and the selected experimental data.

Nickel NP data are reported from experiments that involve the use of nickel NPs as catalyst seeds on top of which carbon nanofibers \cite{luo2015strong} and GaN nanowires \cite{kim2014situ} are grown, at temperatures of $700\text{ K}$ and $750 \text{ K}$, respectively. {For these catalyst-mediated systems, the relevant assumption is that the magnetic field influences the formation and stabilization of the Ni seed size before or during the early stages of catalyst activation, while the subsequent CNF or GaN growth step preserves the field-dependent ordering of the catalyst particle sizes. The comparison therefore tests whether the measured catalyst radius follows the field-selected critical seed scale predicted by the nucleation model.} The radius-field relations for both configurations are shown in Figs. \ref{fig:2}c and \ref{fig:2}d. 

{The results presented in Figs. 2--5 demonstrate the predictability range of our model. The agreement between our model and experiments spans superparamagnetic magnetite and nickel NPs, in addition to diamagnetic silver NPs. This provides a computationally inexpensive alternative to atomistic and electronic-structure simulations. We note that direct atomistic or DFT studies for these exact magnetic-field-assisted synthesis datasets are not available to our knowledge; we therefore cite representative continuum, atomistic, and DFT-based nanomaterial calculations to contextualize computational cost and scope\cite{Giovannini2023-fs,D1RA04876G}.}

{Figure \ref{fig:2} shows that the theory captures the reduction in radius as the magnetic field increases and covers the experimental range almost entirely. Field-resolved parameter sensitivities and comparisons with simple empirical models are provided in Supplementary Section S6, ``Sensitivity Analysis.''} {The distribution-normalized deviations reported in Fig. \ref{fig:2} show the strongest agreement for the Ni systems and indicate that the homogeneous magnetite configuration is captured better than the gradient configuration. In practical terms, $\varepsilon_w$ compares the model-data mismatch with the width of the measured particle-size distribution: smaller values indicate that the predicted mean radius lies closer to the experimental mean relative to the spread of particle sizes. Values below 1 therefore indicate that the model deviation is smaller than the observed distribution width. According to this metric, the best agreement is obtained for Ni-GaN ($\varepsilon_w=0.07$), followed by Ni-CNF ($\varepsilon_w=0.11$), homogeneous magnetite ($\varepsilon_w=0.36$), and gradient-field magnetite ($\varepsilon_w=0.43$). The remaining magnetite deviations are likely related to irregular NP shapes observed in TEM images \cite{ma2022using} and to non-extensive thermodynamic effects that become important at small particle sizes \cite{MANIOTIS2025116285, Guisbiers01012019}.}

Figure \ref{fig:3} shows a comparison between the size distributions predicted by our theory and the experimental distributions for Magnetite NPs. An interesting feature of the model is that it naturally predicts a narrowing of the accessible NP size range as the magnetic field increases. This behavior emerges from the curvature of the free energy landscape and is reflected in the shrinking band of possible radii in Figs.~\ref{fig:2} and \ref{fig:4}. Such behavior is consistent with experimental observations in several magnetic field-assisted synthesis methods and suggests that magnetic fields may provide a practical route for not only controlling the size but also refining the size distributions.

\subsection*{{Susceptibility-Only Limit: Silver NPs}}
\begin{figure}[!ht]
    \centering
    \includegraphics[width=1\linewidth]{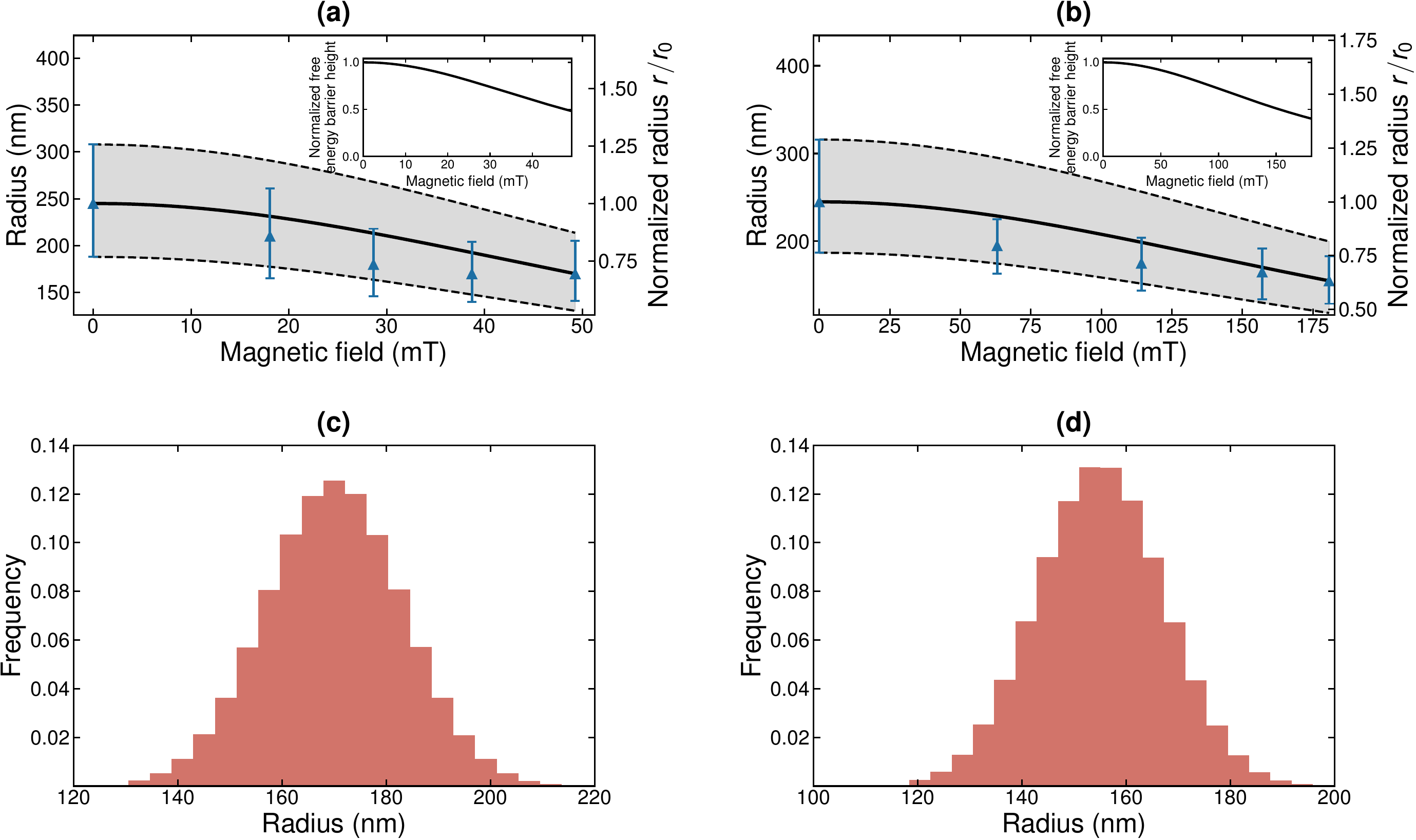}
    \caption{The radius-field relation for Silver NPs \cite{li2025effects} in the (a) parallel and (b) perpendicular configurations. Panels (c) and (d) show the theoretical size distributions at the last experimental point in the parallel and perpendicular configurations, respectively. The blue markers are experimental points, the black line represents the solution of Eq. \eqref{eq:dxdB2} for $x(B=0)=x_0$, the dashed lines are the solutions at the experimental bounds. The shaded region represents the possible sizes predicted by the theory and the insets showcase the effect of the magnetic field on the free energy barrier height.}
    \label{fig:4}
\end{figure}

For NPs that do not possess a permanent magnetic moment, such as Silver NPs \cite{kthiri2021novel,li2025effects}, only the induced magnetization term affects the free energy landscape, so we take the limit $m\to 0$ (or equivalently $\xi\to0$) in Eq. \eqref{eq:dxdB} to obtain: 
\begin{equation}
    \frac{dx}{dB}
=
\frac{2 K B\, n'(x)}
{8\pi \gamma a^2 - \left(\Delta\mu + K B^2\right) n''(x)}
\label{eq:dxdB2}
\end{equation}
{Equation \eqref{eq:dxdB2} is therefore the susceptibility-only limit of the generalized theory. Because this limit contains no logarithmic orientational-entropy contribution from permanent magnetic moments, it is expected to apply to diamagnetic Ag NPs but not necessarily to magnetic NPs whose field response includes moment alignment. Upon integration, Eq. \eqref{eq:dxdB2} exactly recovers our previous analytical description \cite{Tawalbeh2025controlling,Tawalbeh2026}. We first test this limiting case using the Silver NP data of Li et al. \cite{li2025effects}, where the magnetic field is applied parallel or perpendicular to the stirring direction. As shown in Fig. \ref{fig:4}, the reduced equation captures the experimentally observed decrease in Silver NP radius with increasing magnetic field, confirming that the earlier analytical model is recovered as the diamagnetic limit of the present framework.}

Mechanical stirring plays an important role in particle transport and aggregation during the synthesis process, as discussed by Li et al ~\cite{li2025effects}. However, in the present framework, stirring does not directly determine the thermodynamic critical size of the NP. Instead, it primarily acts as a secondary transport mechanism that influences how atoms and clusters are delivered to the nucleating particle. The observed difference between the parallel and perpendicular magnetic field configurations therefore suggests that the magnetic field orientation modifies the effective surface free energy $\gamma$ entering the nucleation free energy. In particular, the parallel configuration appears to correspond to a lower effective surface free energy, which leads to nucleation occurring at a smaller critical radius. Orientation-dependent modifications of surface tension have previously been observed experimentally by Hayakawa et al ~\cite{hayakawa2019effect} and are discussed in detail in our previous work \cite{Tawalbeh2026}.

\begin{figure}[!ht]
    \centering
    \includegraphics[width=1\linewidth]{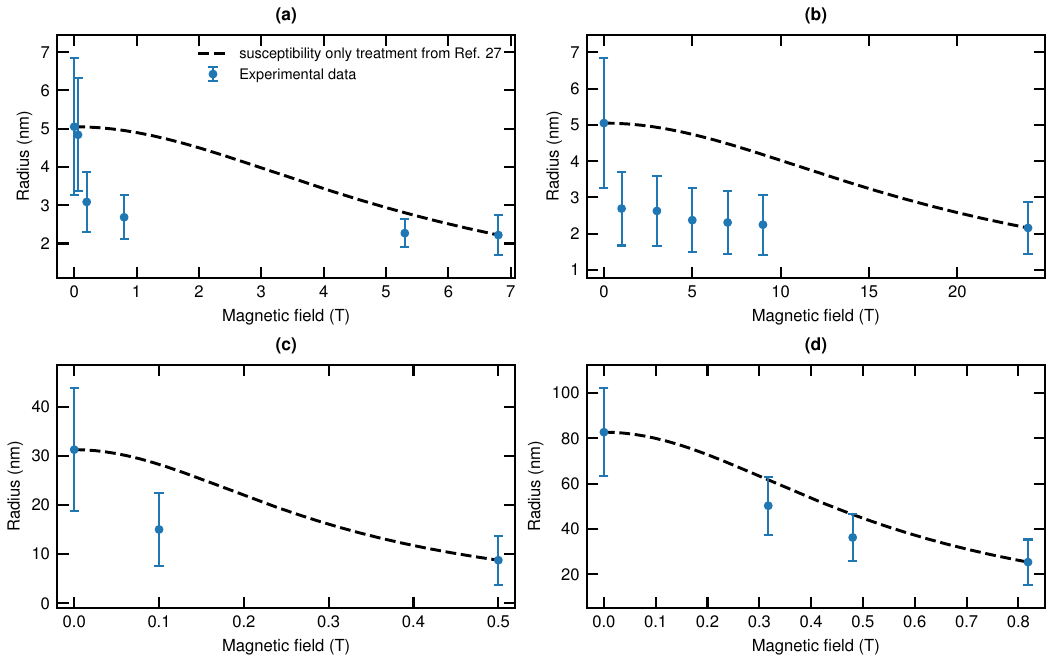}
    \caption{{Comparison between the susceptibility-only limiting equation and experimental radius-field data for (a) magnetite NPs in the gradient configuration, (b) magnetite NPs in the homogeneous configuration, (c) Ni NPs used for carbon nanofiber growth, and (d) Ni NPs used for GaN nanowire growth. The dashed black curves show the reduced analytical model obtained from Eq. \eqref{eq:dxdB2}, while the blue markers show experimental data. The mismatch shows that the limiting equation is not sufficient for magnetic NP systems.}}
    \label{fig:5}
\end{figure}

{To test whether this limiting equation can serve as a general description, we applied the same two-anchor analytical construction to the magnetite and nickel datasets. As shown in Fig. \ref{fig:5}, the susceptibility-only limit captures the endpoint radii by construction but fails to reproduce the intermediate radius-field trends, especially for magnetite and CNF-grown Ni. This comparison demonstrates that the Silver NP result is a valid limiting case, whereas magnetite and nickel require the full formulation in Eq. \eqref{eq:dxdB}, including the logarithmic orientational-entropy term.}

\subsection*{Limitations}
The present model remains a thermodynamic description and therefore neglects kinetic processes such as diffusion-limited growth, aggregation, and coalescence, which may influence NP growth after the initial nucleation stage. Additionally, the sphere-packing representation used for the atomic structure of the NPs is an idealized description that may deviate from the true atomic arrangements in very small clusters. Despite these limitations, the proposed theory is still capable of describing the decreasing trend with larger magnetic field for multiple classes of magnetic materials.

Another limitation arises from the fact that classical thermodynamics assumes extensivity of thermodynamic quantities, an assumption that may break down at nanometer length scales. For very small NPs, the large surface-to-volume ratio can introduce nonextensive effects in which thermodynamic quantities such as the chemical potential or surface free energy become size dependent \cite{MANIOTIS2025116285,Guisbiers01012019}. These effects may modify the detailed shape of the free energy landscape, particularly for clusters containing only a few hundred atoms, and could therefore contribute to the deviations observed in some experimental datasets.
\section*{Materials and Methods}
\subsection*{Theoretical Framework}
We solve Eq.~\eqref{eq:dxdB} numerically as an ordinary differential equation with the initial condition $x(B=0)=x_0$, which we infer from experiments, and treat the parameter set $\boldsymbol{p}=\{\Delta \mu, \gamma, m_0,\delta\}$ as the least-squares fit parameters. We start the solver with a trial parameter set $\boldsymbol{p}_0$ and solve the ODE to obtain the trajectory $x(B;\boldsymbol{p}_0)$. The resulting solution is evaluated at the experimental magnetic field values $B_i$ to produce theoretical radii $r_i^{\mathrm{th}} = a\,x(B_i;\boldsymbol{p}_0)$. These predictions are then compared with the measured radii $r_i^{\mathrm{exp}}$, and the parameters are iteratively updated using a nonlinear least–squares optimization procedure that minimizes the distribution-width-normalized residuals between theory and experiment. The parameter set $\boldsymbol{p}^{\ast}$ that best reproduces the observed radius–field dependence is finally used for the results.

The fitting is performed as a solve-compare-update loop. For each set of trial parameters, the ODE must be integrated again from $B=0$ across the full range of the experimental field, since changing $\Delta\mu$, $\gamma$, $m_0$, or $\delta$ changes the full trajectory. The optimizer therefore does not fit the data in an explicit closed form, but instead searches the parameter space through repeated numerical solutions of Eq.~\eqref{eq:dxdB}. At each iteration, the mismatch between the theoretical and experimental radii is evaluated, and the parameter values are adjusted until the residual no longer decreases appreciably. 

Since the sizes of NPs are often reported as histograms, we use the smallest and largest bins as the size bounds at each magnetic field value. We reevaluate the ODE with the parameter set $\boldsymbol{p}^{\ast}$ at these bounds for $B = 0$ in order to generate a ``band'' of possible sizes according to the predictions of our theory, which is the shaded gray region enclosed by dashed black lines in Figs.~\ref{fig:2}, \ref{fig:4}a and \ref{fig:4}b. At any magnetic field value, we now have three points that represent the theoretical mean NP size (black line) and the theoretical bounds. We use these points to introduce a Gaussian distribution centered on the mean with its tails at the two theoretical bounds, with a standard deviation $\sigma = \min(r_0 - r_{\mathrm{min}},\; r_{\mathrm{max}} - r_0)/3$, placing the bounds at approximately $\pm 3\sigma$. This distribution is then interpreted as the predicted probability density of NP sizes at that magnetic field.

The physical idea behind this procedure is that the ODE describes how a particle of a given initial size evolves under the magnetic field. By solving the same equation for the experimental mean size and for the lower and upper initial bounds, we effectively propagate the full initial size spread through the same thermodynamic evolution law. The spread obtained at the final magnetic field corresponds to the size distribution predicted by the theory. In this sense, the dashed solutions are not error bars, but the field-evolved boundaries of the NP ensemble.

{In the reduced coordinate $x=r/a$, $a$ is not taken as the lattice constant. It is the radius of a sphere with the same volume as one material building unit,}
\[
{
a=\left(\frac{3V_a}{4\pi}\right)^{1/3}.
}
\]
{For an fcc elemental material, the conventional unit cell contains four atoms, so the building-unit volume is $V_a=a_{\mathrm{lat}}^3/4$ rather than $a_{\mathrm{lat}}^3$. For Ni, this gives $V_a=3.52^3/4=10.9~\text{\AA}^3$ and therefore $a=1.38~\text{\AA}$ \cite{haynes2016crc}. For magnetite, the unit-cell volume is $8.396^3\approx592~\text{\AA}^3$; dividing by the eight formula units in the unit cell gives $V_a\approx74.0~\text{\AA}^3$, which gives $a=2.60~\text{\AA}$ \cite{cornell2003iron,haynes2016crc}. These are the values used in the numerical calculations. For silver, we use the effective radius from our previous susceptibility-only treatment.}

To generate the theoretical histograms shown in Figs.~\ref{fig:3} and \ref{fig:4}, we sample $14,000$ times from the Gaussian distribution described above using rejection sampling. First, a candidate radius $r_c$ is drawn uniformly from the interval bounded by the theoretical lower and upper sizes, $[r_{\min},r_{\max}]$. Then a second random number $u$ is drawn uniformly from the interval $[0,1]$. The Gaussian probability density $P(r_c)$ is evaluated at the candidate radius and normalized by its maximum value $P_{\max}$, which occurs at the mean radius. The candidate point is accepted if
\[
u < \frac{P(r_c)}{P_{\max}},
\]
and is rejected otherwise. Repeating this procedure many times produces an ensemble of particle sizes distributed according to the target Gaussian. In other words, points near the theoretical mean are accepted more often, while points near the tails are accepted less often, which produces the intended size distribution. We note that Monte Carlo sampling was used only after the optimal parameter set $\boldsymbol{p}^{\ast}$ had already been determined from the radius–field relation. The fitting is performed on the deterministic ODE solution, while the histogram generation is a second step used only to construct the predicted size distributions from the fitted theory.

To reduce the dependence on any one stochastic realization, the sampling is not performed only once. Instead, the complete histogram-generation procedure is repeated $100$ times with independent random seeds, and the resulting normalized histograms are averaged bin-by-bin before the final renormalization. This reduces Monte Carlo noise and produces a smoother theoretical histogram that is more suitable for comparison with the experimental distributions. The final histogram is binned on a uniform grid chosen to match the approximate bin width used in the corresponding experimental histogram, so that the visual comparison between theory and experiment is done on the same scale.

For materials in the "susceptibility response" limit, which are materials with no permanent magnetic moment, such as silver NPs, we set $m_0=0$, which implies $\xi\to0$. In this limit the logarithmic alignment term associated with permanent magnetic moments vanishes, and Eq.~\eqref{eq:dxdB} reduces to Eq.~\eqref{eq:dxdB2}. The reduced equation depends only on the quadratic susceptibility-response term and can be treated more explicitly. As shown in the Supplementary Information, multiplying Eq.~\eqref{eq:dxdB2} by its denominator leads to an exact differential form, which can then be integrated directly to obtain the first integral of the problem. The resulting integrated expression is
\[
8\pi\gamma a^2 x(B)-\left(\Delta\mu+KB^2\right)n'(x(B))=0,
\]
after imposing the boundary conditions to solve for $\gamma$ and $\Delta\mu$, we obtain exactly the same implicit radius–field relation obtained in our previous work \cite{Tawalbeh2025controlling, Tawalbeh2026}.
\section*{Conclusion}

We have reformulated classical nucleation theory as a geometrically closed thermodynamic framework for magnetic-field-driven NP nucleation. By introducing a sphere-packing representation of atomic assembly, the theory establishes a direct and quantitative link between discrete atomic structure and continuum free-energy contributions, enabling {a quantitative description} of how magnetic fields modify both the nucleation barrier and the critical nucleus size.
The framework provides {a general treatment} of NP nucleation across superparamagnetic, paramagnetic, and diamagnetic systems within a single formalism. It quantitatively reproduces experimental observations for magnetite, nickel, and silver NPs, while {recovering} our previous analytical results as a limiting case of the generalized theory.
Beyond quantitative agreement, the theory reveals that both the reduction in mean particle size and the narrowing of size distributions arise naturally from the geometry of the field-modified free-energy landscape. This identifies magnetic-field-assisted nucleation as a geometrically constrained thermodynamic process rather than a material-specific effect.
{Taken together, these results support a broadly applicable and computationally efficient framework for magnetic-field control of NP size, providing a possible route to model-guided design across distinct magnetic material classes.} {Looking forward, this framework can be used as a design tool for selecting magnetic-field conditions, material parameters, and synthesis windows that target desired NP sizes and size distributions. Because the model connects field strength directly to the nucleation free-energy landscape, it may guide the development of NP systems for catalysis, magnetic materials, electronic nanostructures, and field-assisted nanomanufacturing.}

\section*{Author contributions statement}
M.F.P.\ conceived and coordinated the theoretical framework.  
Y.T.\ derived the generalized thermodynamic theoretical framework, performed the analytical calculations, and generated all the figures.  
All authors contributed to the interpretation of the results and to writing the manuscript.

\section*{Funding}
This publication is based upon work supported by Khalifa University under Award No. CIRA-2021-108.
\section*{Competing interests}
The authors declare no competing interests.
\section*{Data availability}
The datasets used and/or analyzed during the current study are available from the corresponding author on reasonable request.
\section*{Appendix / Supplementary Information}
\renewcommand{\theequation}{S\arabic{equation}}
\setcounter{equation}{0}
\renewcommand{\thefigure}{S\arabic{figure}}
\setcounter{figure}{0}

% --- Tables ---
\renewcommand{\thetable}{S\arabic{table}}
\setcounter{table}{0}
\noindent 
\section*{S1. Statistical alignment contribution for permanent magnetic moments}

The logarithmic contribution to the free energy stems from the orientational statistics of a permanent dipole moment in a thermal bath. This term does not assume $\mathbf{M}\propto\mathbf{B}$ and instead follows from the field–dependent reduction of orientational entropy when a magnetic moment aligns with the field \cite{Pathria2021-nd}.

We consider an effective magnetic moment m associated with the nucleus in a uniform external magnetic field $\mathbf{B}=B\hat{z}$. In the present framework, this moment is taken as $m = n(x)m_0$, proportional to the nanoparticle size, with fixed magnitude for a given nucleus. {The orientational degree of freedom therefore belongs to one effective collective nanoparticle moment, not to $n(x)$ independent microscopic moments.}

The interaction energy of a given orientation is
\begin{equation}
E(\theta) = - m B \cos\theta
\end{equation}
where $\theta$ is the angle between $\mathbf{m}$ and $\mathbf{B}$. The canonical partition function over orientations is
\begin{equation}
Z_{\mathrm{or}}(B) = \int d\Omega \exp\!\left(\beta m B \cos\theta\right)
\end{equation}
with $\beta = 1/(k_B T)$ and $d\Omega = \sin\theta\,d\theta\,d\phi$. Performing the integral gives
\begin{equation}
Z_{\mathrm{or}}(B) = 2\pi\int_{0}^{\pi}\sin\theta\,\exp\!\left(\beta m B \cos\theta\right)d\theta
= 4\pi\,\frac{\sinh\xi}{\xi}
\end{equation}
where the dimensionless field parameter is
\begin{equation}
\xi = \beta m B = \frac{mB}{k_B T}
\end{equation}
The corresponding orientational free energy is
\begin{equation}
F_{\mathrm{or}}(B) = -k_B T \ln Z_{\mathrm{or}}(B)
\end{equation}
Since the factor $4\pi$ is independent of $B$ it can be absorbed into an additive constant that does not affect field–dependent thermodynamic differences. The field–dependent contribution is therefore written as 
\begin{equation}
F_{\log}(B) = -k_B T \ln\!\left(\frac{\sinh\xi}{\xi}\right)
\end{equation}
Next, we add this term to a previously derived free energy expression \cite{Tawalbeh2025controlling, Tawalbeh2026}, which successfully described the formation of diamagnetic Ag NPs under different configurations of an externally applied magnetic field \cite{kthiri2021novel,li2025effects}

\begin{equation}
\Delta F = - n(x)\Delta\mu + \gamma A_s  -
\Theta(\sigma)
\frac{V_{NP}}{2}
\frac{3}{\mu_0}
\frac{|\chi_m|}{3+\chi_m}
B^2 -k_B T \ln\!\left(\frac{\sinh\xi}{\xi}\right)
\end{equation}
Here $\Delta\mu$ is the change in the chemical potential when forming a new phase, $A_s=4\pi r^2$ is the surface energy and $\gamma$ is the surface free energy.$V_{NP}=n(x)V_a$ where $V_a$ is the atomic volume and $\mu_0$ is the magnetic permeability of free space. The orientation factor $\Theta \equiv \left| \langle \cos\theta \rangle \right|
$ allows taking thermal deviations in the magnetization into consideration. We model finite-temperature fluctuation by assuming a Gaussian angular distribution centered at the energetically favored orientation ($\theta_0=0$  for paramagnetism, $\theta_0=\pi$ for diamagnetism), with standard deviation $\sigma$
\begin{equation}
\langle \cos\theta \rangle =
\frac{\int_{0}^{2\pi} \cos\theta \exp\!\left[-\frac{(\theta-\theta_0)^2}{2\sigma^2}\right] d\theta}
{\int_{0}^{2\pi} \exp\!\left[-\frac{(\theta-\theta_0)^2}{2\sigma^2}\right] d\theta}
\end{equation}
The parameter $\sigma$ represents the angular spread induced by thermal agitation and controls the degree of alignment (larger $\sigma$ leads to a larger deviation from the perfect alignment).

{The susceptibility-response term and the orientational-alignment term represent two different magnetic contributions to the free energy. The susceptibility term arises naturally from the response of a magnetizable nucleus to an external magnetic field,}
\begin{equation}
{
F_{\chi}(x,B)=-K n(x)B^2,
}
\end{equation}
{where}
\begin{equation}
{
K=\Theta(\sigma)\frac{V_a}{2}\frac{3}{\mu_0}\frac{|\chi_m|}{3+\chi_m}.
}
\end{equation}
{This term describes how strongly the material responds to the applied field through the susceptibility $\chi_m$ and is linear in the amount of material through $n(x)$. By contrast, the logarithmic term describes the loss of orientational entropy when naturally present magnetic moments become more aligned with the field,}
\begin{equation}
{
F_{\mathrm{or}}(x,B)=-k_BT\ln\!\left[\frac{\sinh\xi}{\xi}\right],
\qquad
\xi=\frac{n(x)m_0B}{k_BT}.
}
\end{equation}
{The two terms can both contain a $B^2$ dependence at low field, but they do not reduce to the same size dependence. Expanding the orientational term for $\xi\ll1$ gives}
\begin{equation}
{
F_{\mathrm{or}}(x,B)
=-k_BT\left(\frac{\xi^2}{6}+O(\xi^4)\right)
=-\frac{[n(x)m_0]^2B^2}{6k_BT}+O(B^4).
}
\end{equation}
{Thus, the two low-field contributions scale differently,}
\begin{equation}
{
F_{\chi}\propto n(x)\chi_m B^2,
\qquad
F_{\mathrm{or}}\propto \frac{n(x)^2m_0^2B^2}{k_BT}.
}
\end{equation}
{This different dependence on particle size, moment scale, and temperature shows that the two terms are not the same mathematical contribution in the present model. We therefore use $\chi_m$ to describe the material response to the field and $m_0$ to describe the magnitude of the naturally present magnetic moment whose orientational entropy changes in the applied field. Ref.~\cite{VanVleck1932} makes the same basic distinction: the magnetic susceptibility describes the response of a material to an applied field, whereas the Langevin-type orientational contribution describes how existing moments become biased toward the field direction. In systems where both effects are relevant, both terms may be included; if a material is treated in the susceptibility-only limit, the orientational contribution is removed by setting $m_0=0$.}

{In applications where $m$ is an effective moment of the nucleus, $m$ is taken as $m=n(x) m_0$ with $m_0$ an atomic or molecular moment, so that $\xi$ becomes size dependent through $n(x)$}
\begin{equation}
\xi(x,B) = \frac{n(x)m_0 B}{k_B T}
\end{equation}
The logarithmic term couples directly to the nucleus size through the atom-count function derived in the upcoming section.

\section*{S2. Atom-count function $n(x)$}

To relate the nanoparticle radius to the atom number, we adopt a core–shell sphere packing model. The interior is assumed to achieve optimal close packing with packing fraction

\[
\phi_b = \frac{\pi}{3\sqrt{2}}
\]
Surface atoms experience reduced packing efficiency due to curvature and incomplete coordination. We therefore introduce a defective surface packing fraction $\phi_d$. Separation of bulk and surface contributions

\begin{equation}
n_{\mathrm{bulk}}(r)
=
\phi_b
\frac{V(r-\delta)}{V_a}
\end{equation}

\begin{equation}
n_{\mathrm{surf}}(r)
=
\phi_d
\frac{V(r) - V(r-\delta)}{V_a}
\end{equation}
Using the reduced coordinate $x=r/a$, where $a$ is the effective atomic radius defined from the atomic volume (formula-unit) $V_a$ in the main text, the atom-count function can be written as follows.

\begin{equation}
n(x)
=
\phi_b (x-\delta)^3
+
\phi_d \left[x^3 - (x-\delta)^3\right]
\end{equation}
The parameter $\phi_d$ is obtained by fitting to the numerical data on the packing of the sphere \cite{pack}. This construction is consistent with the Steiner formula for convex bodies \cite{Morvan2008} and has been introduced in Refs.~\cite{Tawalbeh2025controlling,Tawalbeh2026}.

\section*{S3. Radius–field relation}

The critical nucleus size at a given magnetic field is obtained from the barrier stationarity condition
\begin{equation}
\frac{\partial \Delta F(x,B)}{\partial x} = 0
\end{equation}
In the present framework, the work of formation including magnetic contributions is written as
\begin{equation}
\Delta F(x,B)
=
- n(x)\Delta\mu
+ \gamma A_s(x)
+ F_{\mathrm{mag}}(x,B)
+ F_{\log}(x,B)
\end{equation}
The critical radius $x(B)$ is, therefore, implicitly defined by
\begin{equation}
f(x,B) \equiv \frac{\partial \Delta F}{\partial x} = 0
\end{equation}

Rather than solving this condition independently for each value of $B$, we reformulate the problem as a field-driven evolution along the critical manifold.

\begin{equation}
\frac{d}{dB} f(x(B),B) = 0
\end{equation}
Using the chain rule
\begin{equation}
\frac{\partial^2 \Delta F}{\partial x^2} \frac{dx}{dB}
+
\frac{\partial^2 \Delta F}{\partial x \partial B}
=
0
\end{equation}
which gives the equation
\begin{equation}
\frac{dx}{dB}
=
-
\frac{
\partial^2 \Delta F / \partial x \partial B
}{
\partial^2 \Delta F / \partial x^2
}
\end{equation}
Substituting the quadratic magnetization response term
\begin{equation}
F_{\mathrm{mag}}(x,B)
=
-
\Theta
\frac{n(x)V_a}{2}
\frac{3}{\mu_0}
\frac{|\chi_m|}{3+\chi_m}
B^2
\end{equation}
and the logarithmic alignment term
\begin{equation}
F_{\log}(x,B)
=
{- k_B T \ln\left(\frac{\sinh \xi}{\xi}\right)}
\end{equation}
with
\begin{equation}
\xi(x,B) = \frac{n(x)m_0 B}{k_B T}
\end{equation}
{Using this collective-moment form, the stationarity condition becomes}
\begin{equation}
{
f(x,B)=8\pi\gamma a^2x-n'(x)\left[\Delta\mu+KB^2\right]-n'(x)m_0B L(\xi)=0 .
}
\end{equation}
{The two derivatives entering the critical-manifold evolution are therefore}
\begin{equation}
{
\begin{aligned}
\frac{\partial f}{\partial B}
&=-n'(x)\left[
2KB+m_0L(\xi)+\frac{n(x)m_0^2B}{k_BT}L'(\xi)
\right],\\
\frac{\partial f}{\partial x}
&=8\pi\gamma a^2-n''(x)\left[\Delta\mu+KB^2\right]
-n''(x)m_0B L(\xi)
-\frac{\left[n'(x)\right]^2m_0^2B^2}{k_BT}L'(\xi).
\end{aligned}
}
\end{equation}
This fully specifies the differential equation for $x(B)$
\begin{equation}
{
\frac{dx}{dB}=
\frac{
n'(x)\left[
2KB+m_0 L(\xi)+\frac{n(x)m_0^2B}{k_BT}\left(-\mathrm{csch}^2\xi+\frac{1}{\xi^2}\right)
\right]
}{
8\pi\gamma a^2-n''(x)\left[\Delta\mu+KB^2\right]-n''(x)m_0B L(\xi)-\frac{\left[n'(x)\right]^2m_0^2B^2}{k_BT}\left(-\mathrm{csch}^2\xi+\frac{1}{\xi^2}\right)
}
}
\end{equation}
with $
K
=
\Theta(\sigma)\,
\frac{V_a}{2}\,
\frac{3}{\mu_0}\,
\frac{|\chi_m|}{3+\chi_m}
$, $L(\xi)\equiv \coth\xi-\frac{1}{\xi}$, {$\xi=n(x)m_0B/(k_BT)$, and $-\mathrm{csch}^2\xi+1/\xi^2=dL/d\xi$.}

\section*{S4. Susceptibility Response Limit}

In the susceptibility-response limit $m_0=0$ we take ${\xi\to0}$ and so the differential equation reduces to
\begin{equation}
\frac{dx}{dB}
=
\frac{2 K B\, n'(x)}
{8\pi \gamma a^2 - \left(\Delta\mu + K B^2\right) n''(x)}
\end{equation}
Multiplying by the denominator and rearranging gives an implicit differential form
\begin{equation}
\left[8\pi\gamma a^2-\left(\Delta\mu+K B^2\right)n''(x)\right]dx-2KB\,n'(x)\,dB=0
\end{equation}
We define
\begin{equation}
M(x,B)=8\pi\gamma a^2-\left(\Delta\mu+K B^2\right)n''(x)
\end{equation}
\begin{equation}
N(x,B)=-2KB\,n'(x)
\end{equation}
The exactness condition is $\partial M/\partial B=\partial N/\partial x$. Differentiation gives
\begin{equation}
\frac{\partial M}{\partial B}=-2KB\,n''(x)
\end{equation}
\begin{equation}
\frac{\partial N}{\partial x}=-2KB\,n''(x)
\end{equation}
so the equation is exact. Therefore, there exists a scalar function $I(x,B)$ such that $dI=M\,dx+N\,dB$. Integrating $M$ with respect to $x$ gives
\begin{equation}
I(x,B)=8\pi\gamma a^2x-\left(\Delta\mu+K B^2\right)n'(x)+C(B)
\end{equation}
Taking $\partial/\partial B$ and enforcing $\partial I/\partial B=N$ yields $C'(B)=0$ so $C(B)$ is a constant and the first integral is
\begin{equation}
I(x,B)=8\pi\gamma a^2x-\left(\Delta\mu+K B^2\right)n'(x)=C
\end{equation}
The constant is fixed by the zero-field equilibrium condition at $B=0$. For $x(0)=x_0$ satisfying $\partial\Delta F/\partial x=0$ in the zero field, one has $8\pi\gamma a^2x_0-\Delta\mu\,n'(x_0)=0$ and therefore $C=0$. Hence, the integrated form of the differential equation is
\begin{equation}
8\pi\gamma a^2x(B)-\left(\Delta\mu+K B^2\right)n'(x(B))=0
\end{equation}
which exactly matches the $r$ vs. $B$ relation from our previous work \cite{Tawalbeh2025controlling} (up to the substitution of anchor points).
\section*{S5. Numerical details and histograms generation}
The theoretical size distribution in each applied field is constructed by
propagating the experimentally observed initial distribution forward through $\mathrm{d}r/\mathrm{d}B$, which is numerically integrated using an RK4 scheme from $B = 0$ to the target field $B^{*}$, starting from three distinct initial conditions, (1) the
mean radius $r_{0}$, (2) the upper experimental bound $r_{0} + \sigma_{0}^{+}$, and (3) the
lower experimental bound $r_{0} - \sigma_{0}^{-}$, where $\sigma_{0}^{+}$ and
$\sigma_{0}^{-}$ represent the experimental size distribution bounds at $B = 0$. This yields three trajectories whose endpoints at $B^{*}$ define the predicted mean radius $\mu$, upper limit $r_{\mathrm{max}}$, and lower limit $r_{\mathrm{min}}$ of the size distribution.

The physical interpretation is that each particle in the ensemble follows the same
thermodynamic ODE but from a different starting size; the spread
$[r_{\mathrm{min}},\, r_{\mathrm{max}}]$ at $B^{*}$ therefore represents the
predicted size distribution that results from propagating the initial size distribution
through the field-driven nucleation process. A truncated Gaussian
$\mathcal{N}(r_0,\,\sigma^{2})$ supported on $[r_{\mathrm{min}},\, r_{\mathrm{max}}]$ is then used to model the predicted size distribution at $B^{*}$, where the standard deviation is set to
$\sigma = \min(r_0 - r_{\mathrm{min}},\; r_{\mathrm{max}} - r_0)/3$,
placing the bounds approximately $\pm 3\sigma$ from the mean. To eliminate sensitivity to any single stochastic realization, the procedure is repeated over $N = 100$ independent Monte Carlo draws of $N_{s} = 14{,}000$ samples each using distinct random seeds, and the resulting normalized frequency histograms are averaged bin-by-bin before renormalization. This ensemble averaging reduces Monte Carlo noise
by a factor of $\sqrt{N} = 10$ relative to a single draw, yielding a smooth
theoretical histogram with statistical quality equivalent to $1.4 \times 10^{6}$ samples. The theoretical histogram is binned onto a uniform grid whose bin width matches the mean experimental bin width and whose range spans the full predicted interval $[r_{\mathrm{min}},\, r_{\mathrm{max}}]$.

\section*{{S6. Sensitivity Analysis}}

{To clarify the role of adjustable parameters in the full model, we performed an additional sensitivity and benchmarking analysis using the same audited numerical implementation used to generate Fig. 2 of the main manuscript. The fitted parameter vector is written as $(\Delta\mu,E_s,m_0,\delta)$, where $E_s=4\pi a^2\gamma$. For each dataset, we report field-resolved one-at-a-time percentage radius changes caused by $\pm 20\%$ parameter perturbations and a graphical comparison with simple empirical functions of the magnetic field.}

{The field-resolved sensitivity tables report $100*(r_{\mathrm{pert}}-r_{\mathrm{base}})/r_{\mathrm{base}}$ at each nonzero experimental magnetic-field value. They show that the radius-field trends are primarily controlled by the balance between the bulk driving term $\Delta\mu$ and the collective-moment scale $m_0$, while the predicted radii are comparatively insensitive to changes in $E_s$ and $\delta$ for the fitted parameter sets. The weak local sensitivity to $E_s$ reflects the small fitted surface-energy scale relative to the bulk and magnetic terms in these radius-field fits, especially once the field-dependent terms dominate at higher magnetic fields; it should therefore be interpreted as a local result of the fitted parameter set, not as a general statement that surface energy is unimportant in nucleation. The available experimental datasets are too sparse to provide unique formal uncertainties for all four fitted parameters, especially for the Ni datasets. Therefore, we interpret the fitted parameters as effective material parameters constrained by the measured radius-field trend rather than as independently identifiable microscopic constants.}

\begin{figure}[!ht]
\centering
\includegraphics[width=\textwidth]{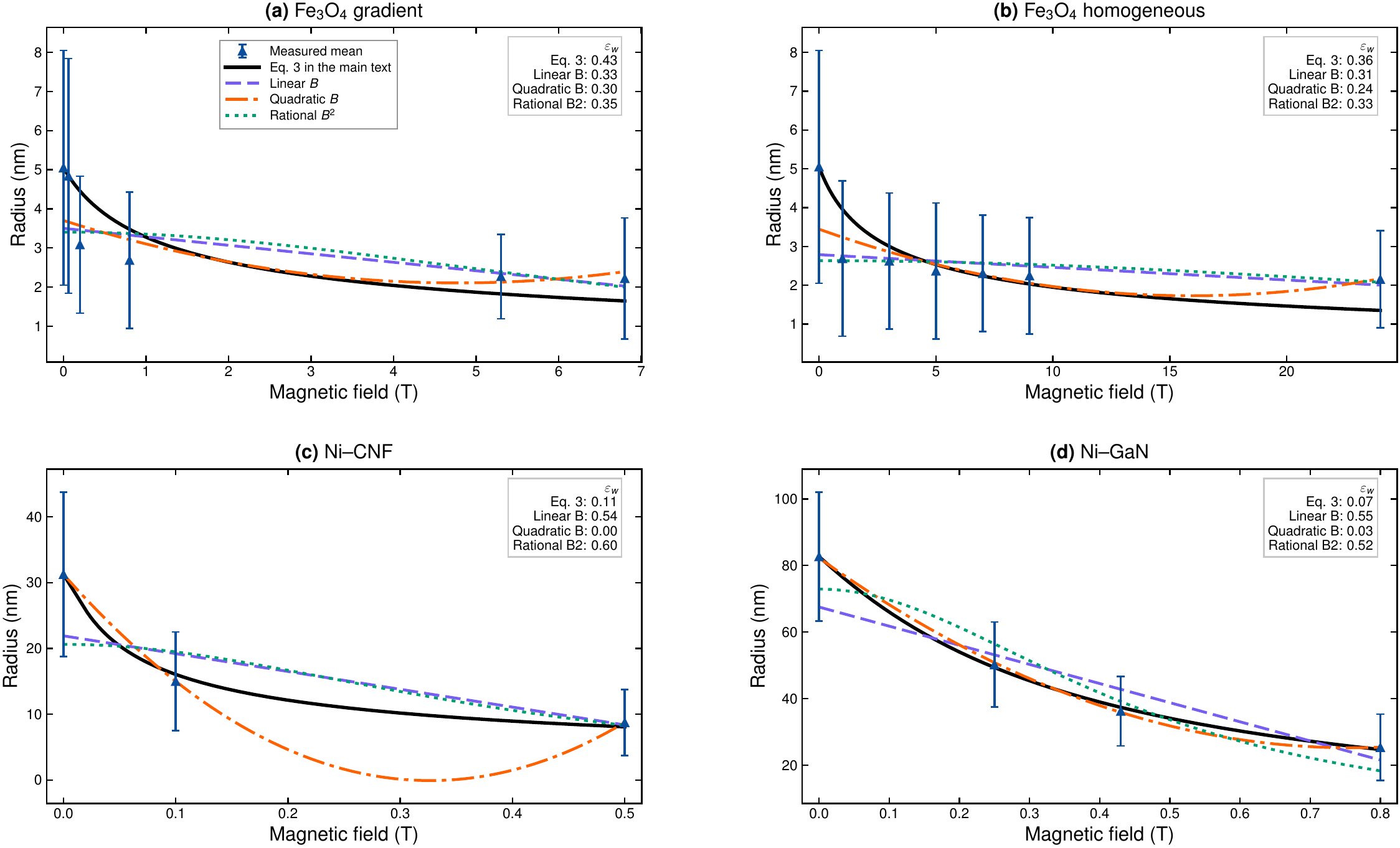}
\caption{{Benchmark comparison between Eq. 3 in the main text and simple empirical functions fitted to the same radius-field datasets. Symbols and curves are defined as follows: blue triangles, measured mean radii with reported distribution widths; black solid line, Eq. 3 in the main text; purple dashed line, linear $B$ fit, $r(B)=c_0+c_1B$; orange dash-dotted line, quadratic $B$ fit, $r(B)=c_0+c_1B+c_2B^2$; green dotted line, rational $B^2$ fit, $r(B)=c_0/(1+c_1B^2)$. The values printed in the upper-right box of each panel are distribution-width-normalized RMS deviations $\varepsilon_w$, where smaller values indicate closer agreement with the measured mean radius relative to the reported particle-size distribution width. The empirical curves provide a check on fitting flexibility: for sparse datasets, especially the Ni--CNF case, a quadratic function can interpolate the data with a very small residual while producing a nonphysical intermediate trend.}}
\label{fig:empirical-benchmark}
\end{figure}

{The empirical-model comparison in Fig. \ref{fig:empirical-benchmark} uses three simple curve shapes as numerical benchmarks. The linear fit, $r(B)=c_0+c_1B$, assumes that the radius changes at a constant rate with magnetic field. The quadratic fit, $r(B)=c_0+c_1B+c_2B^2$, adds curvature and is therefore more flexible, especially for sparse datasets. The rational fit, $r(B)=c_0/(1+c_1B^2)$, represents a monotonic field-dependent decrease that can level off at larger fields. These functions are not nucleation models; they contain no free-energy landscape, no critical-nucleus condition, no distribution-propagation rule, and no connection to the susceptibility-only silver limit. For this reason, we use them only as benchmarks for numerical flexibility rather than as physical replacements for Eq. 3 in the main text.}

\begin{table}[!ht]
\centering
\scriptsize
\caption{{Field-resolved one-at-a-time sensitivity analysis for Fe$_3$O$_4$ gradient. Values are percentage changes in the predicted radius, $100*(r_{\mathrm{pert}}-r_{\mathrm{base}})/r_{\mathrm{base}}$, evaluated at the nonzero experimental magnetic-field values. For each column, only the indicated parameter is perturbed while the remaining fitted parameters and the zero-field radius are held fixed.}}
\label{tab:sensitivity-fe3o4-gradient}
\resizebox{\textwidth}{!}{%
\begin{tabular}{lcccccccc}
\hline
Magnetic field (T) & $\Delta\mu$ $-20\%$ & $\Delta\mu$ $+20\%$ & $E_s$ $-20\%$ & $E_s$ $+20\%$ & $m_0$ $-20\%$ & $m_0$ $+20\%$ & $\delta$ $-20\%$ & $\delta$ $+20\%$ \\
\hline
0.06 & -0.97 & +0.67 & $<0.01$ & $<0.01$ & +0.80 & -0.78 & -0.04 & +0.04 \\
0.2 & -2.63 & +1.88 & $<0.01$ & $<0.01$ & +2.27 & -2.12 & -0.13 & +0.11 \\
0.8 & -5.78 & +4.56 & $<0.01$ & $<0.01$ & +5.55 & -4.71 & -0.40 & +0.33 \\
5.3 & -8.81 & +7.61 & $<0.01$ & $<0.01$ & +9.36 & -7.23 & -0.83 & +0.32 \\
6.8 & -9.09 & +7.87 & $<0.01$ & $<0.01$ & +9.69 & -7.47 & -0.75 & +0.07 \\
\hline
\end{tabular}%
}
\end{table}

\begin{table}[!ht]
\centering
\scriptsize
\caption{{Field-resolved one-at-a-time sensitivity analysis for Fe$_3$O$_4$ homogeneous. Values are percentage changes in the predicted radius, $100*(r_{\mathrm{pert}}-r_{\mathrm{base}})/r_{\mathrm{base}}$, evaluated at the nonzero experimental magnetic-field values. For each column, only the indicated parameter is perturbed while the remaining fitted parameters and the zero-field radius are held fixed.}}
\label{tab:sensitivity-fe3o4-homogeneous}
\resizebox{\textwidth}{!}{%
\begin{tabular}{lcccccccc}
\hline
Magnetic field (T) & $\Delta\mu$ $-20\%$ & $\Delta\mu$ $+20\%$ & $E_s$ $-20\%$ & $E_s$ $+20\%$ & $m_0$ $-20\%$ & $m_0$ $+20\%$ & $\delta$ $-20\%$ & $\delta$ $+20\%$ \\
\hline
1 & -4.40 & +3.32 & $<0.01$ & $<0.01$ & +4.03 & -3.57 & -0.24 & +0.21 \\
3 & -6.90 & +5.64 & $<0.01$ & $<0.01$ & +6.89 & -5.64 & -0.53 & +0.45 \\
5 & -7.77 & +6.54 & $<0.01$ & $<0.01$ & +8.01 & -6.37 & -0.72 & +0.58 \\
7 & -8.21 & +7.01 & $<0.01$ & $<0.01$ & +8.60 & -6.73 & -0.83 & +0.65 \\
9 & -8.48 & +7.30 & $<0.01$ & $<0.01$ & +8.97 & -6.96 & -0.91 & +0.67 \\
24 & -9.38 & +8.18 & $<0.01$ & $<0.01$ & +10.06 & -7.70 & -0.86 & +0.12 \\
\hline
\end{tabular}%
}
\end{table}

\begin{table}[!ht]
\centering
\scriptsize
\caption{{Field-resolved one-at-a-time sensitivity analysis for Ni--CNF. Values are percentage changes in the predicted radius, $100*(r_{\mathrm{pert}}-r_{\mathrm{base}})/r_{\mathrm{base}}$, evaluated at the nonzero experimental magnetic-field values. For each column, only the indicated parameter is perturbed while the remaining fitted parameters and the zero-field radius are held fixed.}}
\label{tab:sensitivity-ni-cnf}
\resizebox{\textwidth}{!}{%
\begin{tabular}{lcccccccc}
\hline
Magnetic field (T) & $\Delta\mu$ $-20\%$ & $\Delta\mu$ $+20\%$ & $E_s$ $-20\%$ & $E_s$ $+20\%$ & $m_0$ $-20\%$ & $m_0$ $+20\%$ & $\delta$ $-20\%$ & $\delta$ $+20\%$ \\
\hline
0.1 & -3.88 & +3.88 & $<0.01$ & $<0.01$ & +4.85 & -3.23 & -0.15 & +0.15 \\
0.5 & -5.43 & +5.50 & $<0.01$ & $<0.01$ & +6.88 & -4.53 & -0.51 & +0.49 \\
\hline
\end{tabular}%
}
\end{table}

\begin{table}[!ht]
\centering
\scriptsize
\caption{{Field-resolved one-at-a-time sensitivity analysis for Ni--GaN. Values are percentage changes in the predicted radius, $100*(r_{\mathrm{pert}}-r_{\mathrm{base}})/r_{\mathrm{base}}$, evaluated at the nonzero experimental magnetic-field values. For each column, only the indicated parameter is perturbed while the remaining fitted parameters and the zero-field radius are held fixed.}}
\label{tab:sensitivity-ni-gan}
\resizebox{\textwidth}{!}{%
\begin{tabular}{lcccccccc}
\hline
Magnetic field (T) & $\Delta\mu$ $-20\%$ & $\Delta\mu$ $+20\%$ & $E_s$ $-20\%$ & $E_s$ $+20\%$ & $m_0$ $-20\%$ & $m_0$ $+20\%$ & $\delta$ $-20\%$ & $\delta$ $+20\%$ \\
\hline
0.25 & -7.16 & +5.82 & $<0.01$ & $<0.01$ & +4.39 & -3.87 & -0.05 & +0.05 \\
0.43 & -8.61 & +7.33 & $<0.01$ & $<0.01$ & +4.29 & -3.80 & -0.09 & +0.09 \\
0.8 & -9.63 & +8.48 & $<0.01$ & $<0.01$ & +3.39 & -3.08 & -0.18 & +0.18 \\
\hline
\end{tabular}%
}
\end{table}

\bibliography{ref}

\end{document}